\documentclass{amsart}
\usepackage{amsfonts}

\newtheorem{theorem}{Theorem}
\theoremstyle{plain}

\newtheorem{corollary}{Corollary}

\newtheorem{definition}{Definition}
\newtheorem{example}{Example}

\newtheorem{lemma}{Lemma}

\newtheorem{proposition}{Proposition}
\newtheorem{remark}{Remark}

\numberwithin{equation}{section}
 
\begin{document}
\title[]{Wigner-Eckart theorem for tensor operators \\
of Hopf algebras}
\author{Marek Mozrzymas}
\address{Institute of Theoretical Physics, University of Wroclaw, \\
pl. Maxa Borna 9, 50-204 Wroclaw, Poland}
\email{marmoz@ift.uni.wroc.pl}
\thanks{}
\date{6 April 2004}
\thanks{}

\begin{abstract}
We prove Wigner-Eckart theorem for the irreducible tensor operators for
arbitrary Hopf algebras, provided that tensor product of their irreducible
representation is completely reducible. The proof is based on the properties
of the irreducible representations of Hopf algebras, in particular on Schur
lemma. Two classes of tensor operators for the Hopf algebra U$_{t}$(su(2))
are considered. The reduced matrix elements for the class of irreducible
tensor operators are calculated. A construction of some elements of the
center of U$_{t}$(su(2)) is given.
\end{abstract}

\maketitle

\section{Introduction}

The concept of tensor operators is very important in applications of
symmetry techniques (Lie groups and algebras) in theoretical physics.
Irreducible tensor operators for the Lie group of space rotations were first
introduced by Wigner \cite{1}. Equivalent definition of tensor operators for
the coresponding Lie algebra was given by Racah \cite{2}. These tensor
operators play very important role in theory of angular momentum in quantum
physics. In the last twenth years a certain classes of Hopf algebras namely
the quantum (deformed) Lie groups and algebras \ have raised a wide interest
among theoretical physicists. This wide interest may be explained by the
fact that quantum Lie groups and algebras are continous deformations of well
known Lie groups and algebras and their representation theory is very
similar to that of nondeformed classical Lie groups and algebras \cite{3}, 
\cite{4}, \cite{5}. In particular it has been shown in several papers \cite%
{6, 7, 8} that for the representations of the algebra $U_{t}(su(2))$ the
Racah-Wigner calculation (i.e. the properties of the irreducible
representations) can be fully developped following the same lines as in the
classical case of $su(2).$

The importance of tensor operators in representation theory of the Lie
groups and algebras leads naturally to investgate the concept of tensor
operator for quantum groups and algebras and in general for Hopf algebras.
The classical Wigner-Racah definition of irreducible tensor operator has
been extented to the quantum lie algebras in papers \cite{7, 8, 9} and
Wigner-Eckart theorem has been proved in the similar way as in classical
undeformed symmetry structures. In papers \cite{10, 11} a new, more general
definitions of tensor operators for arbitrary Hopf algebra has been
proposed. According these definitions tensor operators are homomorphisms \
of some Hopf algebra representations. The new general definitions, on one
hand are equivalent to the classical Wigner-Racah definitions if the
coresponding Hopf algebra is $su(2)$ or $U_{q}(su(2))$, on the other hand
they allows to deduce easier general properties of tensor operators from
basic properties of Hopf algebra representations. In particular in paper 
\cite{11} it has been suggested a possibility of proving Wigner-Eckart
theorem for the Hopf algebras using general properties of operators in
representation spaces of the Hopf algebras. In this paper we follow this
idea.

We use definition of tensor operator for Hopf algebras which is more general
version of definition formulated in paper \cite{11}. Using this definition
we formulate and prove Wigner-Eckart theorem for irreducible tensor
operators for Hopf algebras that tensor product of the irreducible
representation is completely but not necessarily simply reducible. The proof
is based on the properties Hopf algebra representations, in particular a
conclusions from Schur lemma play important role in it. When the tensor
product of irreducible representations is simply reducible, as in case of $%
U_{t}(su(2))$, the Wigner-Eckart theorem takes the conventional form, where
the matrix elements of the components of irreducible tensor operator are
proportional to the Clebsch-Gordan coefficients and the proportionality
coefficient (i.e. the reduced matrix element) is exactly the same
coefficient which appears in Schur lemma. Using the regular and adjoint
representations, as well the general properties of representations of Hopf
algebras we construct two classes of tensor operators for the Hopf algebra $%
U_{t}(su(2)).$ As an application of Wigner-Eckart theorem we calculate the
reduced matrix elements for the class of the irreducible tensor operators.
Finally we give a method of constructing elements of the center of $%
U_{t}(su(2))$. The method is based on properties of the irreducible
representations of the algebra $U_{t}(su(2)).$

This paper has the following structure. Section II contains the basic
definitions and properties \ of Hopf algebra representations, in particular
definition and examples of tensor operators. In section III we consider
Wigner-Eckart theorem. Finally in section IV we consider two classes of
tensor operators for the Hopf algebra $U_{q}(su(2)),$ we calculate the
reduced matrix element the irreducible ones and we give a method of
constructing elements of the center of $U_{t}(su(2))$.

\section{Tensor operators for Hopf algebras}

In this section we introduce our notations and we rewiew the basic
properties of representatios of Hopf algebras. For more details we refer
reader to the book \cite{13}. We begin by recalling the definition of Hopf \
algebra.

\begin{definition}
A Hopf algebra\textrm{\ }is a vector space \textrm{A }over complex field $%
\mathbf{C}$\textrm{\ }with

1) an associative multiplication, $m:\mathrm{A}\otimes \mathrm{A}\rightarrow 
\mathrm{A}$, $m(\mathrm{a}\otimes \mathrm{b})=\mathrm{ab}$, \textrm{a,b}$\in 
$\textrm{A},%
\begin{equation*}
m\circ (id_{\mathrm{A}}\otimes m)=m\circ (m\otimes id_{\mathrm{A}})
\end{equation*}

2) a coassociatve comultiplication, $\Delta :\mathrm{A}\rightarrow \mathrm{A}%
\otimes \mathrm{A}$, $\ \Delta (\mathrm{a})=\sum_{i}\mathrm{a}%
_{i}^{(1)}\otimes \mathrm{b}_{i}^{(2)}$, \textrm{a}$\in $\textrm{A},%
\begin{equation*}
(id_{\mathrm{A}}\otimes \Delta )\circ \Delta =(\Delta \otimes id_{\mathrm{A}%
})\circ \Delta
\end{equation*}

3) a unit, $i:\mathbf{C}\rightarrow \mathrm{A},$%
\begin{equation*}
m\circ (id_{\mathrm{A}}\otimes i)=m\circ (i\otimes id_{\mathrm{A}})=id_{%
\mathrm{A}}
\end{equation*}

4) a counit, $\varepsilon :\mathrm{A}\rightarrow \mathbf{C},$%
\begin{equation*}
(id_{\mathrm{A}}\otimes \varepsilon )\circ \Delta =(\varepsilon \otimes id_{%
\mathrm{A}})\circ \Delta =id_{\mathrm{A}}
\end{equation*}

5) an antipode $S:\mathrm{A}\rightarrow \mathrm{A,}$%
\begin{equation*}
m\circ (id_{\mathrm{A}}\otimes S)\circ \Delta =m\circ (S\otimes id_{\mathrm{A%
}})\circ \Delta =i\circ \varepsilon
\end{equation*}

such that the mappings $\Delta $ and $\varepsilon $ are algebra hommorphisms
i. e.%
\begin{equation*}
\Delta \circ m=(m\otimes m)\circ (id_{\mathrm{A}}\otimes \tau \otimes id_{%
\mathrm{A}})\circ (\Delta \otimes \Delta )
\end{equation*}%
\begin{equation*}
\varepsilon \circ m=\varepsilon \otimes \varepsilon
\end{equation*}

where the map $\tau :\mathrm{A}\otimes \mathrm{A\rightarrow A}\otimes 
\mathrm{A}$ is given by%
\begin{equation*}
\tau (\mathrm{a}\otimes \mathrm{b})=\mathrm{b}\otimes \mathrm{a}
\end{equation*}
\end{definition}

One can show that the antipode $S$ is always an anti-homomorphism of the
algebra and of the coalgebra,%
\begin{equation*}
S(\mathrm{ab})=S(\mathrm{a})S(\mathrm{b}),(S\otimes S)\circ \Delta =\tau
\circ \Delta \circ S.
\end{equation*}%
We will need later on the following identity%
\begin{equation}
\sum_{i,j}(\mathrm{a}_{i}^{(1)})_{j}^{(1)}\otimes S(\mathrm{a}%
_{i}^{(1)})_{j}^{(2)}\mathrm{a}_{i}^{(2)}=\mathrm{a}\otimes \mathbf{1}
\end{equation}%
where \textrm{a}$\in $\textrm{A. }This identity follows from coassociativity
of the coproduct $\Delta .$

For a given Hopf algebra \textrm{A} one can define a left \textrm{A}-module
or, simply an \textrm{A}-module and the coresponding representation as
follows

\begin{definition}
1) A left \textrm{A}-module is a vector space $V$ together with a bilinear
map $(\mathrm{a},v)\rightarrow \mathrm{a}v$ from $\mathrm{A}\times V$ to $V$%
such that 
\begin{equation*}
\forall \mathrm{a,b}\in \mathrm{A}\forall v\in V,\mathrm{a}(\mathrm{b}v)=(%
\mathrm{ab)}v,\mathbf{1}_{\mathrm{A}}v=v.
\end{equation*}%
2) The action of \textrm{A} on an \textrm{A}-module $V$ defines an algebra
homomorphism $\rho $ $:$ \textrm{A} $\rightarrow $ $Hom(V,V)$ by%
\begin{equation*}
\rho (\mathrm{a}).v=\mathrm{a}v
\end{equation*}%
The pair $(V,\rho )$ is called a representation a Hopf algebra \textrm{A}.
The representation $(V,\rho )$ is irreducible if there is no proper subspace 
$V^{^{\prime }}\subset V$ which is invariant under action of the Hopf
algebra \textrm{A} via map $\rho $.
\end{definition}

Conversely any representation $(V,\rho )$ defines the coresponding \textrm{A}%
-module, therefore in the following we will use equivalently the terminology
of modules and representations. Let us recall some examples of Hopf algebra
modules.

\begin{example}
A Hopf algebra \textrm{A} is itself an \textrm{A}-module via the adjoint
action $\rho (\mathrm{a})\equiv ad_{\mathrm{a}}$ given by 
\begin{equation*}
ad_{\mathrm{a}}(\mathrm{b})=\sum_{i}\mathrm{a}_{i}^{(1)}\mathrm{b}S(\mathrm{a%
}_{i}^{(2)}),
\end{equation*}%
for any $\mathrm{a},\mathrm{b}\in \mathrm{A.}$ The adjoint representation
will be denoted $(\mathrm{A},ad)\equiv \mathrm{A}_{ad}.$
\end{example}

\begin{example}
A Hopf algebra $\mathrm{A}$ is a representation space for a left regular
representation $(\mathrm{A},\Lambda )\equiv \mathrm{A}_{\Lambda }$ where%
\begin{equation*}
\Lambda (\mathrm{a}).\mathrm{b}=m(\mathrm{a}\otimes \mathrm{b})=\mathrm{ab}
\end{equation*}%
for any $\mathrm{a},\mathrm{b}\in \mathrm{A.}$
\end{example}

\begin{example}
Let $V$ and $W$ be a modules of Hopf algebra $\mathrm{A}$. The linear space $%
Hom(V,W)$ is an \textrm{A}-module with the action of \textrm{A} on$\ f\in
Hom(V,W)$ defined as follows$\ \ \ \ \ \ \ \ \ \ \ \ \ \ \ \ \ \ \ \ \ \ \ \
\ \ \ \ $%
\begin{equation*}
(\mathrm{a}f)(v)=\sum_{i}\mathrm{a}_{i}^{(1)}fS(\mathrm{a}_{i}^{(2)})(v)
\end{equation*}%
for any $v\in V.$ Equivalently, in language of representations for $(V,\pi
), $ $(W,\rho )$ and $(Hom(V,W),\delta )$ we have%
\begin{equation*}
\delta (\mathrm{a})(f)=\sum_{i}\rho (\mathrm{a}_{i}^{(1)})\circ f\circ \pi
(S(\mathrm{a}_{i}^{(2)})).
\end{equation*}
\end{example}

\begin{example}
The tensor product $V\otimes W$ of two \textrm{A}-modules is an \textrm{A}%
-module by%
\begin{equation*}
\mathrm{a}(v\otimes w)=\sum_{i}\mathrm{a}_{i}^{(1)}v\otimes \mathrm{a}%
_{i}^{(2)}w
\end{equation*}%
for any $v\in V,w\in W$. This yields to the representation $(W\otimes
V,(\rho \otimes \pi )\circ \Delta )$.
\end{example}

The last example is the following

\begin{example}
The counit map $\varepsilon $ of \ $\mathrm{A}$ equips any vector space $V$
with a trivial \textrm{A}-module structure by 
\begin{equation*}
\mathrm{a}v=\varepsilon (\mathrm{a)}v
\end{equation*}%
where $v\in V$ and $\mathrm{a}\in $\textrm{A. }The coresponding
representation $\rho =\varepsilon $ will be called also trivial.
\end{example}

\bigskip The concept of trivial action of the Hopf algebra on vectors of
representation space can be applied to any Hopf algebra representation.

\begin{definition}
\bigskip For any representation $(V,\rho )$ of Hopf algebra $\mathrm{A}$ we
define the subspace of invariant vectors 
\begin{equation*}
V_{\varepsilon }=\{v\in V:\rho (\mathrm{a}).v=\varepsilon (\mathrm{a)}%
v,\forall \mathrm{a}\in \mathrm{A}\}.
\end{equation*}
\end{definition}

\bigskip Next important matematical tool which we are going to use later on
is a homomorphism of representations so let us recall its definition.

\begin{definition}
Let $(V,\rho )$ and $(W,\sigma )$ be representations of the Hopf algebra $%
\mathrm{A}$. A linear map $f\in Hom(V,W)$ is a homomorphism of
representations $(V,\pi )$ and $(W,\rho )$ if%
\begin{equation*}
f\circ \pi (\mathrm{a})=\rho (\mathrm{a})\circ f.
\end{equation*}%
for any \ $\mathrm{a}\in $\textrm{A. }The \ subspace of the algebra
homomorphisms will be denoted $Hom_{\mathrm{A}}(V,W)$.
\end{definition}

\bigskip We give a few examples of homomorphisms of \textrm{A}-modules,
which will be important in the following.

\begin{example}
\bigskip The Hopf algebra $\mathrm{A}$ with the adjoint action $ad_{\mathrm{a%
}}$, $\mathrm{a}\in \mathrm{A}$ form the adjoint representation $(\mathrm{A}%
,ad_{\mathrm{a}})$. On the other hand we have the representation $%
(Hom(V,V),\delta )$ of $\mathrm{A}$ from Example 2. The representation $\rho 
$ $:$ \textrm{A} $\rightarrow $ $Hom(V,V)$ from Definition 2 is a
homomorphism of the Hopf algebra representations.
\end{example}

\begin{example}
A left regular action $\Lambda $ given in Example 2 is a homomorphism of
representations $\mathrm{A}_{ad}$ and ($Hom(\mathrm{A}_{\Lambda },\mathrm{A}%
_{\Lambda }),\delta )$ i.e. $\Lambda \in Hom_{\mathrm{A}}($ $\mathrm{A}%
_{ad},Hom(\mathrm{A}_{\Lambda },\mathrm{A}_{\Lambda })).$ In fact we have 
\begin{equation*}
\Lambda (ad_{\mathrm{a}}(\mathrm{b}))=\sum_{i}\Lambda (\mathrm{a}%
_{i}^{(1)})\Lambda (\mathrm{b)}\Lambda (S(\mathrm{a}_{i}^{(2)}))
\end{equation*}%
for any $\mathrm{a},\mathrm{b}\in \mathrm{A}$ which means that 
\begin{equation*}
\Lambda \circ ad_{\mathrm{a}}=\delta (\mathrm{a})\circ \Lambda .
\end{equation*}
\end{example}

\bigskip Let introduce some useful notations and mappings. If $f\in Hom(V,W)$
where $(V,\pi )$ and $(W,\rho )$ are representations of the Hopf algebra $%
\mathrm{A}$ then we define%
\begin{equation}
\pi _{f}(\mathrm{a})\equiv f\circ \pi (\mathrm{a}):V\rightarrow W
\end{equation}%
and the linear mapping $m_{\rho }^{\pi }:Hom(W)\otimes Hom(V,W)\rightarrow
Hom(V,W)$ is defined in the following way%
\begin{equation}
m_{\rho }^{\pi }(\rho (\mathrm{a})\otimes \pi _{f}(\mathrm{b}))=(m_{\rho
}^{\pi }\circ (\rho \otimes \pi _{f})).(\mathrm{a}\otimes \mathrm{b})\equiv
\rho (\mathrm{a})\circ \pi _{f}(\mathrm{b}).
\end{equation}

The following theorem describes the relation between the homomorphisms of
representations and invariant linear mappings

\begin{theorem}
\bigskip Let $(V,\pi )$ and $(W,\rho )$ be representations of the Hopf
algebra $\mathrm{A}$. The linear mapping $f\in Hom(V,W)$ is a homomorphism
of representations if and only if $f$ is an invariant element of the
representation $(Hom(V,W),\delta )$. That is we have for any \ $\mathrm{a}%
\in $\textrm{A} 
\begin{equation*}
f\circ \pi (\mathrm{a})=\rho (\mathrm{a})\circ f\Longleftrightarrow \delta (%
\mathrm{a})(f)=\sum_{i}\rho (\mathrm{a}_{i}^{(1)})\circ f\circ \pi (S(%
\mathrm{a}_{i}^{(2)}))=\varepsilon (\mathrm{a)}f
\end{equation*}%
or equivalently we have $Hom_{\mathrm{A}}(V,W)=(Hom(V,W))_{\varepsilon }$.
\end{theorem}

\begin{remark}
\bigskip When the Hopf algebra is the enveloping algebra $\mathrm{U(L)}$ of
a Lie algebra \textrm{L }with the standard Hopf algebra structure $\Delta (%
\mathrm{x})=\mathrm{x}\otimes \mathbf{1}+\mathbf{1}\otimes \mathrm{x},$ $%
\varepsilon (\mathrm{x)}=0$ if $\mathrm{x}$ $\neq \mathbf{1}$, $\varepsilon (%
\mathbf{1})=1$, $S(\mathrm{x})=-\mathrm{x}$ for \textrm{x}$\in \mathrm{L}$,
the two equations in the Theorem 1 coincident for \textrm{x}$\in \mathrm{L}$%
, so in the case of the Lie algebra $\mathrm{L}$ the theorem is trivial.
\end{remark}

\begin{corollary}
\bigskip An element $\mathrm{a}$ of the Hopf algebra $\mathrm{A}$ is $ad$%
-invariant if and only if $\mathrm{a}$ belongs to the center $Z(\mathrm{A})$
of $\mathrm{A}$ i.e. $\mathrm{A}_{\varepsilon }=Z(\mathrm{A}).$
\end{corollary}

We will later on need Schur lemma which we present in the following form

\begin{lemma}
\bigskip Let $(V,\pi )$ and $(W,\rho )$ be irreducible finite dimensional
representations of the Hopf algebra $\mathrm{A}$ and \bigskip $f\in Hom_{%
\mathrm{A}}(V,W)$ then $f$ is either isomorphism of $V$ on $W$ and $f=\alpha
id_{V}$ $(\alpha \in \mathrm{k})$ if $V=W$, or $f=0$.
\end{lemma}

\bigskip Using Schur lemma one can prove

\begin{proposition}
\bigskip \bigskip \bigskip Let $(V,\pi )$ and $(W,\rho )$ be finite
dimensional representations of the Hopf algebra $\mathrm{A}$ such that 
\begin{equation*}
V\otimes W=\oplus _{i}U_{i}
\end{equation*}%
where $(U_{i},\sigma _{i})$ are irreducible representations of the Hopf
algebra $\mathrm{A}$ and let \bigskip $f\in Hom_{\mathrm{A}}(V\otimes W,U)$
where $(U,\sigma )$ is also an irreducible representation $\mathrm{A}.$ Then 
\begin{equation*}
f=\oplus _{i}f_{i}
\end{equation*}%
where $f_{i}=f\mid _{U_{i}}:U_{i}\rightarrow U$ and $f_{i}$ is either an
isomorphism of $U_{i}$ on $U$ and $f=\alpha _{i}id_{U_{i}}$ $(\alpha _{i}\in 
\mathrm{k})$ if $U_{i}=U$, or $f=0$.
\end{proposition}

The above proposition follows directly from the fact that \bigskip $f_{i}\in
Hom_{\mathrm{A}}(U_{i},U)$ and from Schur lemma.

\bigskip Before formulation a corollary from the Proposition 1 which will be
important later on we give some remarks concerning the representations that
we will have to deal with. We will consider irreducible finite dimensional
representations of the Hopf algebra $\mathrm{A}$ $(V,\pi )$, $(W,\rho )$
such that their tensor product $V\otimes W$ is completely reducible but not
necessarily simply reducible i.e. 
\begin{equation*}
V\otimes W=\oplus _{K=1}^{M}(\oplus _{i=1}^{I_{K}}W^{K}(i))
\end{equation*}%
where $(W^{K}(i),\rho ^{K})$ are irreducible representations of $\mathrm{A}$
such that $W^{K}(i)$ and $W^{K^{^{\prime }}}(i^{^{\prime }})$ are not $%
\mathrm{A}$- isomorphic if $K\neq K^{^{\prime }}$ and $I_{K}$ express how
many times the irreducible representation $W^{K}$ appears in the direct sum.
Let $\{e_{m}^{V}\}$, $\{e_{n}^{W}\}$ and $\{e_{l}^{K}(i)\}$ be the basis of
respectively $W$, $W$ and $W^{K}(i)$ then for any $e_{m}^{V}\in V$, $%
e_{n}^{W}\in W$ we have Glebsch-Gordan decomposition 
\begin{equation*}
e_{m}^{V}\otimes e_{n}^{W}=\sum_{K=1}^{M}\sum_{i=1}^{I_{K}}\sum_{l}\lambda
_{l}^{K}(i;m,n)e_{l}^{K}(i).
\end{equation*}

\bigskip Now we can formulate a corollary form the Proposition 1.

\begin{corollary}
Let $(V,\pi )$, $(W,\rho )$ and $(W^{K}(i),\rho ^{K})$ be irreducible finite
dimensional representations of the Hopf algebra $\mathrm{A}$ such as
described above i.e.%
\begin{equation*}
V\otimes W=\oplus _{K=1}^{M}(\oplus _{i=1}^{I_{K}}W^{K}(i))
\end{equation*}%
where $W^{K}(i)$ and $W^{K^{^{\prime }}}(i^{^{\prime }})$ are not $\mathrm{A}
$- isomorphic if $K\neq K^{^{\prime }}$ ,$W^{K}(i)=W^{K}(i^{^{\prime }})$
for $i,i^{^{\prime }}=1,...I_{K}$ and for any basis vectors $e_{m}^{V}\in V$%
, $e_{n}^{W}\in W$ we have 
\begin{equation*}
e_{m}^{V}\otimes e_{n}^{W}=\sum_{K=1}^{M}\sum_{i=1}^{I_{K}}\sum_{l}\lambda
_{l}^{K}(i;m,n)e_{l}^{K}(i)
\end{equation*}%
and let \bigskip $f\in Hom_{\mathrm{A}}(V\otimes W,W^{L})$ where $L=1,...,M$
then for any $e_{m}^{V}\in V$, $e_{n}^{W}\in W$ 
\begin{equation*}
f(e_{m}^{V}\otimes e_{n}^{W})=\sum_{i=1}^{I_{L}}\alpha _{i}\sum_{l}\lambda
_{l}^{L}(i;m,n)e_{l}^{L}
\end{equation*}%
where $\alpha _{i}\in \mathbf{C}$ and $\alpha _{i}$ do not depend on the
indices $m,n$ of the bases $\{e_{m}^{V}\}$, $\{e_{n}^{W}\}$ and $%
\{e_{l}^{L}\}$.
\end{corollary}

\begin{proof}
\bigskip From the Proposition 1 and under assumptions on $W^{K}(i)$ we have 
\begin{equation*}
f=\oplus _{K=1}^{M}(\oplus _{i=1}^{I_{K}}f^{K}(i))
\end{equation*}%
and%
\begin{equation*}
f^{K}(i)=f\mid _{W^{K}(i)}:W^{K}(i)\rightarrow W^{L},f^{K}(i)=\alpha
_{i}\delta ^{KL}id_{W^{L}}.
\end{equation*}%
Therefore%
\begin{eqnarray*}
f(e_{m}^{V}\otimes e_{n}^{W}) &=&[\oplus _{K=1}^{M}(\oplus
_{i=1}^{I_{K}}f^{K}(i))].(\sum_{K=1}^{M}\sum_{i=1}^{I_{K}}\sum_{l}\lambda
_{l}^{K}(i;m,n)e_{l}^{K}(i)) \\
&=&\sum_{i=1}^{I_{L}}\alpha _{i}\sum_{l}\lambda _{l}^{L}(i;m,n)e_{l}^{L}.
\end{eqnarray*}%
and it is clear that $\alpha _{i}$ do not depend on the indices $m,n$ of the
bases $\{e_{m}^{V}\}$, $\{e_{n}^{W}\}$ $and$ $\{e_{l}^{L}\}.$
\end{proof}

Now we give the definition of tensor operators for Hopf algebras .

\begin{definition}
Let $(V,\pi )$, $(W,\rho )$ and $(U,\sigma )$ be representations of the Hopf
algebra $\mathrm{A}$ and let $\mathit{T}\in Hom(V,Hom(W,U))$ then $\mathit{T}
$ is a tensor operator of type $V$ in $W$ if $\mathit{T}\in Hom_{\mathrm{A}%
}(V,Hom(W,U))$ . In other words tensor operator $\mathit{T}$ is a
homomorphism of \ \textrm{A}-modules $V$ and $Hom(W,U)$ and it satisfies 
\begin{equation}
\mathit{T}\circ \pi (\mathrm{a})=\delta (\mathrm{a})\circ \mathit{T}
\end{equation}%
where $\delta $ is the action of the Hopf algebra $\mathrm{A}$ in the
representation $(Hom(W,U),\delta )$.

Let vectors $\{e_{l}^{V}\}_{l\in I\subset N}$ be a basis of the
representation space $V$, then the linear operators $\mathit{T}%
(e_{l}^{V})\equiv \mathit{T}_{l}^{V}\in Hom(W,U)$ will be called the
components of the tensor operator $\mathit{T}$. If $\dim V<\infty $ then the
components $\mathit{T}_{i}^{V}$ of $\mathit{T}$ satisfie%
\begin{equation}
\pi (\mathrm{a})_{jl}\mathit{T}_{j}^{V}=\sum_{i}\sigma (\mathrm{a}%
_{i}^{(1)})\circ \mathit{T}_{l}^{V}\circ \rho (S(\mathrm{a}_{i}^{(2)}))
\end{equation}%
where $\pi (\mathrm{a})_{jl}$ is a matrix of $\pi (\mathrm{a})$. If all the
representations $(V,\pi )$, $(W,\rho )$ and $(U,\sigma )$ are irreducibles
then the tensor operator $\mathit{T}$ is called irreducible.
\end{definition}

\begin{remark}
\bigskip \bigskip This definition is more general version of the definition
given in paper \cite{11} where $W=U$.
\end{remark}

\bigskip From Theorem 1 it follows immediately

\begin{proposition}
$\mathit{T}\in Hom(V,Hom(W,U))$ is a tensor operator if and only if $\mathit{%
T}\in (Hom(V,Hom(W,U)))_{\varepsilon }$ i.e. if and only if 
\begin{equation*}
\varepsilon (\mathrm{a})\mathit{T}=\sum_{i}\delta (\mathrm{a}%
_{i}^{(1)})\circ \mathit{T}\circ \pi (S(\mathrm{a}_{i}^{(2)})).
\end{equation*}
\end{proposition}

The above definition of tensor operator although seems to be abstract is a
generalisation of the classical definitions of tensor operators given by
Wigner and Racah \cite{1, 2}. It can be seen when the Hopf algebra $\mathrm{A%
}$ is an enveloping algebra $\mathrm{U(L)}$ of a Lie algebra \textrm{L }with
the standard Hopf algebra structure $\Delta (\mathrm{x})=\mathrm{x}\otimes 
\mathbf{1}+\mathbf{1}\otimes \mathrm{x},$ $\varepsilon (\mathrm{x)}=0$ if $%
\mathrm{x}$ $\neq \mathbf{1}$, $\varepsilon (\mathbf{1})=1$, $S(\mathrm{x})=-%
\mathrm{x}$ for \textrm{x}$\in \mathrm{L}$, and the representation space $V$
is finite dimensional spanned by a fixed basis $\{e_{i}\}_{i=1...n}$ where
operators $\pi (\mathrm{x})$ are represented by matrices $\pi (\mathrm{x}%
)_{ij}.$Then the components of a tensor operators $\mathit{T}_{i}$ satisfie
the equation%
\begin{equation*}
\pi (\mathrm{x})_{ij}\mathit{T}_{i}=\delta (\mathrm{x})(\mathit{T}%
_{j})=\sigma (\mathrm{x})\circ \mathit{T}_{j}-\mathit{T}_{j}\circ \rho (%
\mathrm{x})
\end{equation*}%
where \textrm{x}$\in \mathrm{L}$ and in particular if $(W,\rho )\equiv $ $%
(U,\sigma )$ it can be written in the form%
\begin{equation*}
\pi (\mathrm{x})_{ij}\mathit{T}_{i}=[\rho (\mathrm{x}),\mathit{T}_{j}]
\end{equation*}%
which is precisely Racah definition of components of tensor operator in $V.$
Similarly if the Hopf algebra $\mathrm{A}$ is a group algebra $k\mathrm{G}$
of the group $\mathrm{G}$\textrm{\ }with the Hopf algebra structure $\Delta (%
\mathrm{g})=\mathrm{g}\otimes \mathrm{g},$ $\varepsilon (\mathrm{g)}=1$, $S(%
\mathrm{g})=\mathrm{g}^{-1}$ for \textrm{g}$\in \mathrm{G}$, and the
representation space $V$ is finite dimensional spanned by a fixed basis $%
\{e_{i}\}_{i=1...n}$ where operators $\pi (\mathrm{g})$ are represented by
matrices $\pi (\mathrm{g})_{ij}.$In this case the components of a tensor
operators $\mathit{T}_{i}$ satisfie the equation%
\begin{equation*}
\pi (\mathrm{g})_{ij}\mathit{T}_{i}=\delta (\mathrm{g})(\mathit{T}%
_{j})=\sigma (\mathrm{g})\circ \mathit{T}_{j}\circ \rho (\mathrm{g}^{-1})
\end{equation*}%
which is the same equation as in the Wigner definition when $(W,\rho )\equiv 
$ $(U,\sigma ).$

Let us give some examples of tensor operators.

\begin{example}
The Example 6 shows that the representatoin $\rho $ from Definition 2 is
itself a tensor operator because $\rho \in Hom_{\mathrm{A}}(\mathrm{A}%
,Hom(W\otimes W)).$
\end{example}

\begin{example}
The left regular action $\Lambda $ of $\mathrm{A}$ om itself as defined in
Example 2 is a tensor operator because $\Lambda \in Hom_{\mathrm{A}}($ $%
\mathrm{A}_{ad},Hom(\mathrm{A}_{\Lambda },\mathrm{A}_{\Lambda }))$ (Example
7).
\end{example}

\section{\protect\bigskip Wigner-Eckart theorem for Hopf algebras}

In this section we shall formulate and prove Wigner-Eckart theorem for
irreducible tensor operators. Before presentation of the proof we have to
formulate the following lemma which will be important in the proof of the
Wigner-Eckart theorem.

\begin{lemma}
Assume that

1) $(V,\pi ),W,\rho ),$ $(U,\sigma )$ and $(Hom(W,U),\delta )$ are
representations of the Hopf algebra $\mathrm{A,}$

2) $\mathit{T}\in Hom(V,Hom(W,U)),$

3) $\mathit{\check{T}}\in Hom(V\otimes W,U))$ and $\check{T}(v\otimes w)=%
\mathit{T}(v).w$ \ $\forall v\in V,w\in W.$ \ 

\ Then we have:\ \ \ \ \ \ \ \ \ \ \ \ \ \ \ \ \ \ \ \ \ \ \ \ \ \ \ \ \ \ \
\ \ \ \ \ \ \ \ \ \ \ \ \ \ \ \ \ \ \ \ \ \ \ \ \ \ \ \ \ \ \ \ \ \ \ \ \ \
\ \ \ \ \ 

a) $\mathit{T}\in Hom_{\mathrm{A}}(V,Hom(W,U))$\ \ \ \ i.e. $\forall \mathrm{%
a\in A}$ $\mathit{T}\circ \pi (\mathrm{a})=\delta (\mathrm{a})\circ \mathit{T%
}$ \ \qquad\ \ \ \ 

if and only if\ \ \ \ \ \ \ \ \ \ \ \ \ \ \ \ \ \ \ \ \ \ \ \ \ \ \ \ \ \ \
\ \ \ \ \ \ \ \ \ \ \ \ \ \ \ \ \ \ \ \ \ \ \ \ \ \ \ \ \ \ \ \ \ \ \ \ \ \
\ \ \ \ \ \ \ \ \ \ \ \ \ \ \ \ \ \ \ \ \ \ \ \ \ \ \ \ \ \ \ \ \ \ \ \ \ \
\ \ \ \ \ 

b) $\mathit{\check{T}}\in Hom_{\mathrm{A}}(V\otimes W,U))$ \ i.e. $\forall 
\mathrm{a\in A}$ $\mathit{\check{T}}\circ \lbrack (\pi \otimes \rho )\Delta (%
\mathrm{a})]=\sigma (\mathrm{a})\circ \mathit{\check{T}}$.
\end{lemma}

This lemma has been formulated in \cite{11} in a little less general form
where $(W,\rho )\equiv (U,\sigma )$, therefore we will prove this lemma.

\begin{proof}
Let us prove $(\Rightarrow ).$ The action $\delta $ of representation $%
(Hom(W,U),\delta )$ is given in Example 2. We rewrite the condition a) for $%
\mathit{T}$ in the form%
\begin{equation}
\mathit{T}[\pi (\mathrm{a}).v].w=\{\sum_{i}\sigma (\mathrm{a}%
_{i}^{(1)})\circ \mathit{T}(v)\circ \rho (S(\mathrm{a}_{i}^{(2)}))\}.w
\end{equation}%
for any $\mathrm{a\in A}$, $v\in V$, $w\in W.$ We have to prove that from
this it follows the condition b) for $\mathit{\check{T}}$ which can be
written as follows%
\begin{equation*}
\sum_{i}\mathit{T}[\pi (\mathrm{a}_{i}^{1}).v].(\rho (\mathrm{a}%
_{i}^{(2)}).w)=\sigma (\mathrm{a})[\mathit{T}(v).w]
\end{equation*}%
for any $\mathrm{a\in A}$, $v\in V$, $w\in W.$ Applying in LHS of the above
equation condition (3.1) for $\mathrm{a}=\mathrm{a}_{i}^{1}$\ we get 
\begin{equation*}
\sum_{i}\mathit{T}[\pi (\mathrm{a}_{i}^{1}).v].(\rho (\mathrm{a}%
_{i}^{(2)}).w)=\sum_{ij}\{\sigma \lbrack (\mathrm{a}_{i}^{(1)})_{j}^{(1)}]%
\circ \mathit{T}(v)\circ \rho \lbrack S(\mathrm{a}_{i}^{(1)})_{j}^{(2)}]\}%
\circ \rho (\mathrm{a}_{i}^{(2)}).w
\end{equation*}%
In the notation (2.2), (2.3) it takes the form 
\begin{equation*}
\sum_{i}\mathit{T}[\pi (\mathrm{a}_{i}^{1}).v].(\rho (\mathrm{a}%
_{i}^{(2)}).w)=\{m_{\sigma }^{\rho }\circ (\sigma \otimes \rho _{\mathit{T}%
(v)}).(\sum_{ij}(\mathrm{a}_{i}^{(1)})_{j}^{(1)}\otimes S(\mathrm{a}%
_{i}^{(1)})_{j}^{(2)}\mathrm{a}_{i}^{(2)})\}.w
\end{equation*}%
Using the identity (2.1) we get 
\begin{equation*}
\sum_{i}\mathit{T}[\pi (\mathrm{a}_{i}^{1}).v].(\rho (\mathrm{a}%
_{i}^{(2)}).w)=\{m_{\sigma }^{\rho }\circ (\sigma \otimes \rho _{\mathit{T}%
(v)}).(\mathrm{a}\otimes \mathbf{1})\}.w=\sigma (\mathrm{a})[\mathit{T}%
(v).w].
\end{equation*}%
Similarly one can prove the implication $(\Leftarrow ).$
\end{proof}

Now we can formulate Wigner-Eckart theorem for irreducible tensor operators.

\begin{theorem}
Let $(V,\pi )$, $(W,\rho )$ and $(W^{K}(i),\rho ^{K})$ be irreducible finite
dimensional representations of the Hopf algebra $\mathrm{A}$ such that 
\begin{equation*}
V\otimes W=\oplus _{K=1}^{M}(\oplus _{i=1}^{I_{K}}W^{K}(i))
\end{equation*}%
where $W^{K}(i)$ and $W^{K^{^{\prime }}}(i^{^{\prime }})$ are not $\mathrm{A}
$- isomorphic if $K\neq K^{^{\prime }}$, $W^{K}(i)=W^{K}(i^{^{\prime }})$
for $i,i^{^{\prime }}=1,...I_{K}$ and for basis vectors $e_{m}^{V}\in V$, $%
e_{n}^{W}\in W$ we have Clebsch-Gordan decomposition%
\begin{equation*}
e_{m}^{V}\otimes e_{n}^{W}=\sum_{K=1}^{M}\sum_{i=1}^{I_{K}}\sum_{l}\lambda
_{l}^{K}(i;m,n)e_{l}^{K}(i)
\end{equation*}%
where $\{e_{m}^{V}\},$ $\{e_{n}^{W}\}$ and $\{e_{l}^{K}(i)\}$ are bases of $%
V,$ $W$ and $W^{K}(i)$ respectively and let 
\begin{equation*}
\mathit{T}\in Hom_{\mathrm{A}}(V,Hom(W,W^{L}))
\end{equation*}%
where $L=1,...,M$ be a tensor operator. \ \ 

Then we have the following formula for matrix elements of the components \ $%
\mathit{T}(e_{m}^{V})$ of the tensor operator $\mathit{T}$%
\begin{equation*}
\mathit{T}(e_{m}^{V})_{\ln }=\sum_{i=1}^{I_{L}}\alpha _{i}\lambda
_{l}^{L}(i;m,n).
\end{equation*}%
where $\alpha _{i}\in \mathbf{C}$ and $\alpha _{i}$ do not depend on $l,m,n.$
\end{theorem}

\begin{proof}
From Lemma 2 we know that linear mapping $\mathit{\check{T}}\in Hom(V\otimes
W,W^{L}))$, $L=1,...,M$%
\begin{equation*}
\check{T}(e_{m}^{V}\otimes e_{n}^{W})=\mathit{T}(e_{m}^{V}).e_{n}^{W}
\end{equation*}%
is a homomorphism of representations. Then from Corollary 2 we get%
\begin{equation*}
\mathit{T}(e_{m}^{V}).e_{n}^{W}=\mathit{\check{T}}(e_{m}^{V}\otimes
e_{n}^{W})=\sum_{l}\sum_{i=1}^{I_{L}}\alpha _{i}\lambda
_{l}^{L}(i;m,n)e_{l}^{L}.
\end{equation*}%
On the other hand the matrix of the operator $\mathit{T}(e_{m}^{V})$ is
defined by equation%
\begin{equation*}
\mathit{T}(e_{m}^{V}).e_{n}^{W}=\sum_{l}\mathit{T}(e_{m}^{v})_{\ln }e_{l}^{L}
\end{equation*}%
Comparing two last equations we get the statement of the theorem.
\end{proof}

According to the Wigner-Eckart theorem the matrix elements of the components
of the irreducible tensor operators are linear combinations of coefficients
of the Clebsch-Gordan serie. Therefore if we know all the coefficients $%
\lambda _{l}^{L}(i;m,n)$ and if know also $I_{L}$ particular values of
matrix elements of irreducible tensor components $\mathit{T}(e_{m}^{V})$
then we can calculate all $I_{L}$ ceoffcients $\alpha _{i}$ and than express
all remaining matrix elements of $\mathit{T}(e_{m}^{V})$ by coefficients $%
\lambda _{l}^{L}(i;m,n)$. The simplest \ case is when the tensor product $%
V\otimes W$ is simply reducible i.e. when each representation $W^{K}$
appears in the Clebsch-Gordan serie only once. Then the matrix elements of $%
\mathit{T}(e_{m}^{V})$ are proportional to Clebsch-Gordan coefficients $%
\lambda _{l}^{L}(i;m,n).$ This case will be discussed in the next section.

\section{Applications of Wigner-Eckart theorem to the quantum algebra $%
U_{t}[su(2)]$}

Wigner-Eckart theorem formulated in the previous section can be applied to a
wide class of Hopf algebras. In particular it can be applied to
Drinfeld-Jimbo deformations i.e. the quantum deformations $U_{t}(L)$ of the
enveloping algebras of complex simple Lie algebras \cite{3}, \cite{12}
because any finite dimensional representation of $U_{t}(L)$ is completely
reducible so the assumptions of the Wigner-Eckart theorem are satisfied.

In this section we will consider irreducible tensor operators for the
quantum algebra $U_{t}[sl(2)]$ which is generated by elements $e,$ $f,$ $%
k^{\pm 1}$ with relations%
\begin{equation*}
kk^{-1}=k^{-1}k=\mathbf{1;}ke=t^{2}ek;kf=t^{-2}fk
\end{equation*}%
and%
\begin{equation*}
\lbrack e,f]=\frac{k^{2}-k^{-2}}{t^{2}-t^{-2}}.
\end{equation*}%
The coproduct structure is the following%
\begin{equation*}
\Delta (k^{\pm 1})=k^{\pm 1}\otimes k^{\pm 1};\Delta (e)=e\otimes
k^{-1}+k\otimes e;\Delta (f)=f\otimes k^{-1}+k\otimes f
\end{equation*}%
\begin{equation*}
\varepsilon (k^{\pm 1})=1;\varepsilon (e)=\varepsilon (f)=0,
\end{equation*}
the antipode is defined by%
\begin{equation*}
S(k)=k^{-1};S(e)=-t^{-2}e;S(f)=-t^{2}f
\end{equation*}%
and $t\in \mathbf{C.}$

If the parameter $t$ is real then one can introduce in $U_{t}[sl(2)]$ an
anti-involution $\ast $ 
\begin{equation*}
e^{\ast }=f;f^{\ast }=e;k^{\ast }=k.
\end{equation*}%
A real form of $U_{t}[sl(2)]$ coresponding to the anti-involution $\ast $ is
denoted $U_{t}[su(2)]$ and it may be considered as deformation of a real Lie
algebra $su(2).$

All irreducible representations of $U_{t}[su(2)]$ are finite dimensional of
highest weight. The irreducible representations $(V^{l},\pi ^{l})$ of $%
U_{t}[su(2)]$ is parametrized by integer or half-integer $l$ and $\dim
V^{l}=2l+1.$The generators $e,f,k^{\pm 1}$ in a weight basis $\{e_{m}^{l}\}$%
, $m=l,l-1,...,-l$ of $V^{l}$ in the following way%
\begin{equation}
\pi ^{l}(e).e_{m}^{l}=([l-m][l+m+1])^{\frac{1}{2}}e_{m+1}^{l}
\end{equation}%
\begin{equation}
\pi ^{l}(f).e_{m}^{l}=([l+m][l-m+1])^{\frac{1}{2}}e_{m-1}^{l}
\end{equation}%
\begin{equation}
\pi ^{l}(k^{\pm 1}).e_{m}^{l}=t^{\pm 2m}e_{m}^{l}
\end{equation}%
where $[n]=\frac{t^{2n}-t^{-2n}}{t^{2}-t^{-2}}.$ In the following we will
use only weight bases in $V^{l}.$

Now we construct in $U_{t}[su(2)]$ irreducible representations of highest
weight $l$ which will be irreducible subrepresentations of adjoint
representation $(U_{t}[su(2)],ad)$.

\begin{proposition}
Let us define for any natural $l$ 
\begin{equation*}
\lambda _{m}^{l}=\left( \frac{[l+m]!}{[2l]![l-m]!}\right) ^{\frac{1}{2}%
}adf^{l-m}.e^{l}k^{-l}
\end{equation*}%
where $m=l,l-1,...,-l,$ then 
\begin{equation}
ade.\lambda _{m}^{l}=([l-m][l+m+1])^{\frac{1}{2}}\lambda _{m+1}^{l}
\end{equation}%
\begin{equation}
adf.\lambda _{m}^{l}=([l+m][l-m+1])^{\frac{1}{2}}\lambda _{m-1}^{l}
\end{equation}%
\begin{equation}
adk^{\pm 1}.\lambda _{m}^{l}=t^{\pm 2m}\lambda _{m}^{l}
\end{equation}%
Therefore the vectors $\lambda _{m}^{l}$ form a basis of a finite
dimensional irreducible representation $(U^{l},ad)\equiv U_{ad}^{l}$ of $%
U_{t}[su(2)]$ where $U^{l}\subset U_{t}[su(2)].$
\end{proposition}

\begin{proof}
A direct calculation shows that $\lambda _{l}^{l}$ is a highest weight
vector of weight $t^{2l}$. The applying the standard procedure of
construction of irreducible highest weight modul of $U_{t}[su(2)]$ we get
the result.
\end{proof}

\begin{corollary}
\bigskip The elements $\lambda _{m}^{l}\in U_{t}[su(2)]$ are components of
the tensor operator $\Lambda ^{l}\in Hom_{U_{t}[su(2)]}($ $%
U_{ad}^{l},Hom(U_{t}[su(2)]_{\Lambda },U_{t}[su(2)]_{\Lambda })).$ According
to Definition 5 it is not an irreducible tensor operator.

\begin{proof}
The left regular action $\Lambda :U_{t}[su(2)]_{ad}\rightarrow
Hom(U_{t}[su(2)]_{\Lambda },U_{t}[su(2)]_{\Lambda })$ is a tensor operator
(Example 9) and $U^{l}$ is an irreducible submodul of $U_{t}[su(2)].$ So it
is obvious that $\Lambda ^{l}:U_{ad}^{l}\rightarrow
Hom(U_{t}[su(2)]_{\Lambda },U_{t}[su(2)]_{\Lambda })$ is also a tensor
operator. \ The equations (4.4-6) show that the components $\lambda _{m}^{l}$
of $\Lambda ^{l}$ satisfie the defining equation (2.5).
\end{proof}
\end{corollary}

Tensor product of two irreducible representation $(V^{k},\pi ^{k})$ and $%
(V^{l},\pi ^{l})$ is completely and simply reducible i.e. we have 
\begin{equation*}
V^{k}\otimes V^{l}=\oplus _{j=\mid k-l\mid }^{k+l}V^{j}
\end{equation*}%
and for the basis vectors $\{e_{m}^{l}\}$ and $\{e_{n}^{j}\}$ we have%
\begin{equation*}
e_{p}^{k}\otimes e_{n}^{l}=\sum_{j=\mid k-l\mid }^{k+l}(kp,jn\mid
jm)_{t}e_{m}^{j}.
\end{equation*}%
Where the coefficients $(kp,jn\mid jm)_{t}$ are Clebsch-Gordan coefficients
(C-Gc). These coefficients has been calculated using various methods and
their analytical formulae can be found in \cite{6, 7, 8}. Thus irreducible
representations of $U_{t}[su(2)]$ satisfie the assumptions of the
Wigner-Eckart theorem formulated in the previous section and in the case of $%
U_{t}[su(2)]$ this theorem takes the form

\begin{theorem}
If $\mathit{T}\in Hom_{U_{t}[su(2)]}(V^{k},Hom(V^{l},V^{j}))$ is an
irreducible tensor operator then the matrix elements of its components $%
\mathit{T}(e_{p}^{k})$ are proprtional to Clebsch-Gordan coefficients i.e.%
\begin{equation*}
\lbrack \mathit{T}(e_{p}^{k})]_{mn}=\alpha (kp,\ln \mid jm)_{t}
\end{equation*}%
where $\alpha $ is a real number called reduced matrix element which do not
depend on $p,m,n.$
\end{theorem}

This result has been obtained earlier in papers \cite{7, 8, 9}. Let us
consider now a tensor operator $\mathit{T}\in
Hom_{U_{t}[su(2)]}(U^{l},Hom(V^{k},V^{j}))$ where $(U^{l},ad)$ is the
irreducible representation given in Proposition 3. From Wigner-Eckart
theorem it follows that matrix elements of components $\mathit{T}(\lambda
_{m}^{l})_{pq}\equiv $ $(\mathit{T}_{m}^{l})_{pq}$ ($\{\lambda _{m}^{l}\}$
is a basis of $U^{l}$ ) are proportional to C-Gc 
\begin{equation*}
(\mathit{T}_{m}^{l})_{pq}=\alpha (lm,kq\mid jp)_{t}
\end{equation*}%
Thus it is sufficient to know one particular value of matrix element $(%
\mathit{T}_{m}^{l})_{pq}$ to determine the reduced matrix element $\alpha $
and then to express all remaining matrix elements $(\mathit{T}_{m}^{l})_{pq}$
in terms of C-Gc. We will calculate the reduced matrix element $\alpha $ in
case when $\mathit{T}=\pi ^{j}$ (i.e. when $\mathit{T}$ is a reperestation
(see Example 6, 8)) and $V^{k}=V^{j}$ in other words we will calculate the
matrices $\pi ^{j}(\lambda _{m}^{l})_{pn}\equiv \lbrack \lambda
_{m}^{l}(j)]_{pn}$ of the basis vectors $\lambda _{m}^{l}$ of $(U^{l},ad)$
in the representation $(V^{j},\pi ^{j}).$Using the defining commutation
relations for $U_{t}[sl(2)]$ one can show that $\lambda _{m}^{l}$ are rather
complicated combination of elements $e,f,k^{\pm 1}$%
\begin{eqnarray}
\lambda _{m}^{l} &=&\left( \frac{[l+m]!}{[2l]![l-m]!}\right) ^{\frac{1}{2}%
}\sum_{p=0}^{N}(-1)^{p}\frac{[l]![l-m]!}{[p]![l-p]!}e^{l-p}f^{l-m-p}\times 
\notag \\
&&\times \left( \sum_{i=0}^{l-m-p}\frac{(-1)^{i}t^{2i}t^{2(l+m)i}}{%
[i]![l-m-i-p]!}\frac{[k+m+p-i-1]!}{[k+m-i-1]!}\right) k^{-m}.
\end{eqnarray}%
where $N=\min (l,l-m)$ we use a symbolic notation 
\begin{equation*}
\lbrack k+m]\equiv \frac{k^{2}t^{2m}-k^{-2}t^{-2m}}{t^{2}-t^{-2}}
\end{equation*}%
and 
\begin{equation*}
\frac{\lbrack k+m+p]!}{[k+m]!}\equiv \lbrack k+m+p]...[k+m+1].
\end{equation*}%
So a direct calculation of $\pi ^{j}(\lambda _{m}^{l})_{pn}$ using matrices $%
\pi ^{j}(e)_{mn}$, $\pi ^{j}(e)_{mn},$ $\pi ^{j}(k^{\pm 1})_{mn}$ seems to
be difficult in general case. However due to Wigner-Eckart theorem it is not
necessary to do it. In fact we have

\begin{proposition}
The basis vectors $\lambda _{m}^{l}$ of $(U^{l},ad)$ have the following
matrix form in the irreducible representation $(V^{j},\pi ^{j})$ 
\begin{equation*}
\pi ^{j}(\lambda _{p}^{l})_{mn}=\alpha (lp,jn\mid jm)_{t}
\end{equation*}%
where 
\begin{equation*}
\alpha =t^{l(l+1)}[l]!\left( \frac{[2j+l+1]!}{[2l]![2j-l]![2j+1]!}\right) ^{%
\frac{1}{2}}
\end{equation*}%
is a reduced matrix element of the irreducible tensor operator $\pi
^{j}:U^{l}\rightarrow Hom(V^{j},V^{j}).$
\end{proposition}

\begin{proof}
The representation $\pi ^{j}:U_{t}[su(2)]\rightarrow Hom(V^{j},V^{j})$ is
itself a tensor operator. Because $U^{l}$ is an irreducible submodul of $%
U_{t}[su(2)]$ then $\pi ^{j}:U^{l}\rightarrow Hom(V^{j},V^{j})$ is an
irreducible tensor opereator. Thus according to the Wigner-Eckart theorem we
have the following expression for matrix element of components $\pi
^{j}(\lambda _{p}^{l})$ of $\pi ^{j}$ 
\begin{equation*}
\pi ^{j}(\lambda _{p}^{l})_{mn}=\alpha (lp,jn\mid jm)_{t}
\end{equation*}%
and in particular 
\begin{equation}
\pi ^{j}(\lambda _{l}^{l})_{mn}=\alpha (ll,jn\mid jm)_{t}.
\end{equation}%
Now on one hand from (4.1-3) we have 
\begin{equation*}
\pi ^{j}(\lambda _{l}^{l})_{mn}=\left( \frac{[j-m+l]![j+m]!}{[j+m-l]![j-m]!}%
\right) ^{\frac{1}{2}}t^{-2l(m-l)}\delta _{mn+l}
\end{equation*}%
and on the other hand we have \cite{8}%
\begin{eqnarray*}
(ll,jn &\mid &jm)_{t}=t^{-\{-l(l+1)+2(m+1)l\}}\times \\
&&\times \left( \lbrack 2j+1]\frac{[2l]![2j-l]![j+m]![j-m+l]!}{%
[2j+l+1]![l]![l]![j-m]![j+m-l]!}\right) ^{\frac{1}{2}}\delta _{mn+l}.
\end{eqnarray*}%
After substitution of two last equations to equation (4.8 ) we get the value
of $\alpha .$
\end{proof}

Let us consider briefly the properties of components $\pi ^{j}(\lambda
_{m}^{l})\equiv \lbrack \lambda _{m}^{l}(j)]$ of the irreducible tensor
operator $\pi ^{j}:U^{l}\rightarrow Hom(V^{j},V^{j}).$ It satisfies the
equation (see Definition 5)%
\begin{equation*}
\pi ^{l}(\mathrm{a})_{nm}\lambda _{n}^{l}(j)=\sum_{i}\pi ^{j}(\mathrm{a}%
_{i}^{(1)})\circ \lambda _{m}^{l}(j)\circ \pi ^{j}(S(\mathrm{a}_{i}^{(2)}))
\end{equation*}%
where $\mathrm{a}\in U_{t}[su(2)]$. In particular for $\mathrm{a}=e,f$ it
takes the form%
\begin{equation*}
([l-m][l+m+1])^{\frac{1}{2}}\lambda _{m+1}^{l}(j)=\pi ^{j}(e)\circ \lambda
_{m}^{l}(j)\circ \pi ^{j}(k^{-1})-t^{-2}\pi ^{j}(k)\circ \lambda
_{m}^{l}(j)\circ \pi ^{j}(e)
\end{equation*}%
\begin{equation*}
([l+m][l-m+1])^{\frac{1}{2}}\lambda _{m-1}^{l}(j)=\pi ^{j}(f)\circ \lambda
_{m}^{l}(j)\circ \pi ^{j}(k^{-1})-t^{2}\pi ^{j}(k)\circ \lambda
_{m}^{l}(j)\circ \pi ^{j}(f)
\end{equation*}%
And it is clear that in the limit $t\rightarrow 1$ if we set formally $%
k=t^{H}$ where $H=[e,f]$ we obtain the classical Racah definition of $su(2)$
irreducible tensor operator.

Let us observe that using a particular C-Gc $(jm,\ln \mid 00)_{t}$ and the
explicite formula (4.7) for $\lambda _{m}^{l}$ we can construct elements of
the center of the algebra $U_{t}[su(2)].$ In fact it is known that the
particular C-Gc 
\begin{equation*}
(jm,in\mid 00)_{t}=\frac{(-1)^{j-m}t^{2m}}{\sqrt{[2j+1]}}\delta _{ji}\delta
_{m,-n}
\end{equation*}%
couple two arbitrary irreducible representations $(V^{j},\pi ^{j})$ and $%
(V^{i},\pi ^{i})$ to one-dimensional trivial represntation. Therefore for
any two irreducible representations $(U^{j},ad)$ and $(U^{i},ad)$ the
following element $\mathfrak{C}$ of $U_{t}[su(2)]$%
\begin{equation*}
\mathfrak{C=}\sum_{mn}(jm,in\mid 00)_{t}\lambda _{m}^{j}\lambda _{n}^{i}
\end{equation*}%
form one dimensional trivial representation $(\mathfrak{C},\varepsilon )$
described in Example 5. It means that $\mathfrak{C}\in
(U_{t}[su(2)])_{\varepsilon }$ and consequently from Corollary 1 it belongs
to the center of $U_{t}[su(2)]$. A particular case 
\begin{equation*}
\mathfrak{C=}\sum_{mn}(1m,1n\mid 00)_{t}\lambda _{m}^{1}\lambda _{n}^{1}
\end{equation*}%
has been given in \cite{11}.

\end{document}